# Partially-Precise Computing Paradigm for Efficient Hardware Implementation of Application-Specific Embedded Systems

Mohsen Faryabi, Amir Hossein Moradi, and Hamid Reza Mahdiani

*Abstract*— Nowadays, the number of emerging embedded systems rapidly grows in many application domains, due to recent advances in artificial intelligence and internet of things. The main inherent specification of these application-specific systems is that they have not a general nature and are basically developed to only perform a particular task and therefore, deal only with a limited and predefined range of custom input values. Despite this significant feature, these emerging applications are still conventionally implemented using general-purpose and precise digital computational blocks, which are essentially developed to provide the correct result for all possible input values. This highly degrades the physical properties of these applications while does not improve their functionality. To resolve this conflict, a novel computational paradigm named as partially-precise computing is introduced in this paper, based on an inspiration from the brain information reduction hypothesis as a tenet of neuroscience. The main specification of a Partially-Precise Computational (PPC) block is that it provides the precise result only for a desired, limited, and predefined set of input values. This relaxes its internal structure which results in improved physical properties with respect to a conventional precise block. The PPC blocks improve the implementation costs of the embedded applications, with a negligible or even without any output quality degradation with respect to the conventional implementation. The applicability and efficiency of the first instances of PPC adders and multipliers in a Gaussian denoising filter, an image blending and a face recognition neural network are demonstrated by means of a wide range of simulation and synthesis results.

*Index Terms*—Bioinspiration, computational blocks, computing paradigm, imprecise computing approximate computing, partially-precise computing, information reduction hypothesis.

## I. INTRODUCTION

T he advent and fusion of some research topics including but not limited to artificial intelligence, video, image, and voice processing, and internet of things, should be enumarated as the main enabler for development of many emerging embedded systems. These application-specific systems are basically developed to perform only a limited set of specified tasks. Meanwhile they are also highly sensitive to system price (in terms of chip area or the number of transistors) based on their widespread utilization by worldwide customers. Moreover, these applications should also deliver exceeded norms in terms of power consumption and delay due to their mobile and real-time usage respectively. These requirements are technically translated into some rigid area, power, and delay constraints, which harden their design and implementation process. Despite their tight physical property constraints, these emerging applications are conventionally realized in digital technology by utilizing universal precise computational blocks (e.g. adders and multipliers), just like any traditional application. The main important specification of the conventional precise computational blocks is that they are developed in an application independent manner. Therefore, they can be generally utilized in all applications without any reservation. However, this also implies that as much as necessary extra cost and performance overheads in terms of area, power, and delay must be paid when designing a precise computational block, to exhaustively preserve its correct output for all possible input combinations and maintain its versatility.

A neglected major conflict just arises here when a universal precise computational block is utilized to implement a custom application that has not a general nature and is inherently developed to perform a particular task and therefore, deals with only a limited and predictable range of input values. In this case, generality of the exploited precise computational blocks highly degrades system physical properties which limits its mobility and real-time performance, while does not improve its functionality. To resolve this significant inconsistency with conventional digital precise computing, the biological source as well as the corresponding novel bio-inspired computing paradigm are introduced in this paper. The computational blocks that are developed based on this paradigm are designed in an application oriented manner, customarily based on the particular requirements of a specific application.

The brain as the most intelligent existing entity in the world has always been an important inspiration source for development of many engineering ideas at different abstraction levels from algorithm to circuit [1]-[7]. As a plausible instance, the imprecise or approximate computing [1], [8] is a promising computational level paradigm which is inspired from the internal brain operation in transferring the

<sup>H</sup>amid Reza Mahdiani and Mohsen Faryabi, are with Department of Computer Science and Engineering, Shahid Beheshti University, Tehran, Iran. (email: mahdiani@gmail.com; m_faryabi@sbu.ac.ir)

Hamid Reza Mahdiani is with Institute of Medical Science and Technology, Shahid Beheshti University, Tehran, Iran

Amirhossein Moradi was with Department of Computer Science and Engineering, Shahid Beheshti University, Tehran, Iran. He is now with the Department of Computer Science and Engineering, Sharif University of Technology. (email: am.moradi@sharif.edu)

values and performing the calculations by means of electro-chemical reactions. In this paper, the well-known information reduction hypothesis about the relation between the brain and its sensing peripherals is introduced as the source of a novel computational level inspiration. This inspiration significantly improves the implementation costs of the computational blocks hardware utilized in custom and application-specific embedded systems, just similar to the imprecise computing. A detailed technical comparative study between the specifications of the imprecise and partially-precise computing paradigms is beyond the scope of this paper. However, the experiments demonstrate that as both of these paradigms are inspired from the biological brain operation, they are entirely compatible just similar to their biological counterparts. Therefore, they can be utilized in a complementary manner to intensify their advantages with respect to their individual usage improve the results.

The next section first briefly explains about the information reduction hypothesis as the source of inspiration. The Partially-Precise Computing paradigm and its underlying basic issues are then introduced next. Section III provides a straightforward design flow which can be utilized for development of the customized partially precise computational blocks of all types for any specific application. The next three sections provide a wide range of experimental results about development, usage, and advantages of customized partially precise adders and multipliers for the Gaussian denoising filter, image blending and face recognition neural network applications. The last section concludes the paper.

## II. BIO-INSPIRED PARTIALLY-PRECISE COMPUTING PARADIGM

According to the information reduction [9] or dimensionality reduction [10] hypothesis as a tenet of sensory neuroscience, each surviving creature learns to distinguish between "task-relevant" and "task-redundant" information according to its vital requirements. It then alleviates its brain tasks by focusing only on processing of the task-relevant information [11]. Experimentally, the task-redundant information is perceptually pruned before entering the brain conceptual processing [9]. Moreover, the task-redundant information once identified, is even no longer perceived and processed within the creature's cognitive system [9]. According to this hypothesis, while there exist infinite amounts of visual or auditory information in a natural environment, creatures of different types see and hear differently when they are looking at the same natural scene or listening to the same natural sound. Undoubtedly, what a creature captures from the environment by means of its perceptual equipment, is its own task-relevant information which tightly depends on its vital requirements [12] [13]. Some in depth explanations about the customized visual system of the human and animals as an interesting instance of the information reduction is provided in the supplementary section I.

The significant engineering inspiration that follows the information reduction hypothesis is that as a custom embedded application essentially deals with a limited and predefined range of task-relevant input values, it is not mandatory to implement it by means of conventional precise computational blocks, which are designed with many overheads in order to accept all (i.e. task-relevant and task-redundant) input values. Based on this approach, a novel bio-inspired computational paradigm called partially-precise computing or sparse computing is introduced in this paper. According to this paradigm, the precise computation result is required only for a limited and predefined subset of task-relevant sparse input values. This relaxes the design process of the corresponding computational block which might result in various benefits in terms of improved complexity, performance, physical properties, and so on. As an instance in digital VLSI territory, a designer might utilize some unusual adders and multipliers with intentionally relaxed structures, to provide the correct result only for a limited set of input values. By assigning a Don't-Care (DC) to the output results of the omitted task-redundant inputs, the Truth-Table (TT) and therefore the internal structure of a bio-inspired Partially-Precise Computational (PPC) block is simplified to provide improved cost and performance. The achieved PPC block loses its generality and cannot be universally utilized in all applications. In a more technical description, partially precise computing trades the computational block generality to achieve useful gains, including but not limited to better performance and physical properties. The good news is that as the sparsity might either "naturally" exist or "intentionally" be inserted on the inputs of a specific application, these improvements are achieved either "without any" or "with some adjustable" amount of system output quality degradation with respect to a conventional implementation.

To better illustrate the operation mechanism of the partially precise computing, it is useful to compare it against well-known imprecise computing paradigm. The imprecise computing improves the system physical properties by inserting the imprecision inside the system which relaxes the internal structure of its computational blocks. On the other hand, the partially precise computing improves the system physical properties by inserting the imprecision on the primary inputs of the system which simplifies its computational blocks. The definition as well as the main characteristics of the two significant sources of sparsity (i.e. natural and intentional) and their effects on development of efficient PPC blocks are explained in the following.

### A. Natural Sparsity: Specifications and Advantages

There are many application-specific embedded systems in which, some levels of sparsity naturally exist on their primary inputs. This sparsity can be exploited for development of customized PPC blocks with better cost/performance for each specific application. There are some significant points about the PPC blocks developed based on natural sparsity. The first and the most important point is that utilization of these blocks improve the system's physical properties without any system output accuracy degradation with respect to the conventional precise implementation. The next important point is that naturally existing sparsity on primary inputs of a system might propagate to the inputs of the deeper computational blocks in the system through first level computational blocks. This also lets to replace the next level conventional blocks in the system with their PPC counterparts. And finally, it is also important to note that the sparsity might be naturally created on the inputs of the deeper computational blocks, even when there is no natural sparsity on the system primary inputs.

*B. Intentional Sparsity: Preprocessing Concept*

Beside some probable natural sparsity which might inherently exist on the system primary inputs, in some applications it is also possible to intentionally eliminate some less significant details of the input based on application vital requirements (such as eliminating the image background in a face recognition application as will be explained). This further increases the input sparsity and results in more improved PPC blocks. The elimination process is applied to the system primary inputs before entering the system (just similar to its biological inspiration source) by means of some simple operations called preprocessing. A preprocessing is a relatively simple, zero or low-cost operation. It is customarily developed based on inherent features of a specific application to increase sparsity on the block inputs as more as possible in a manner to not degrade overall system output quality below an acceptable threshold. The amount of intentionally created sparsity by a preprocessing is controlled by means of one or more preprocessing parameter(s). Utilization of a preprocessing might introduce some controllable and negligible enough deviation in the system output accuracy with respect to the conventional implementation. However, it also highly improves the physical properties of the system PPC blocks. The Down-Sampling and Thresholding are the two first instances of the custom preprocessings as introduced in the following.

*1) Down-Sampling Preprocessing*

The Down-Sampling with the parameter 'x' ($DS_x$) is a preprocessing which increases sparsity by mapping each 'x' consecutive input values to a similar value, while 'x' is a power of 2. For an N-bit unsigned input whose values are in the range of {0 to $2^n - 1$}, the $DS_x$ maps each input value 'i' to its neighbor value $i - (i\ MOD\ x)$. To achieve the PPC blocks with the least cost and good enough precision for a specific application, the maximum down-sampling parameter should be chosen based on the tolerance of that specific application against sparsity. Figs. 1(a)-(d) demonstrate the normalized Gaussian histogram of an image and its preprocessed versions with $DS_2$, $DS_4$, and $DS_8$ respectively. As it is clear, applying the $DS_x$ preprocessing to a signal decreases the number of values by a factor of 1/x.

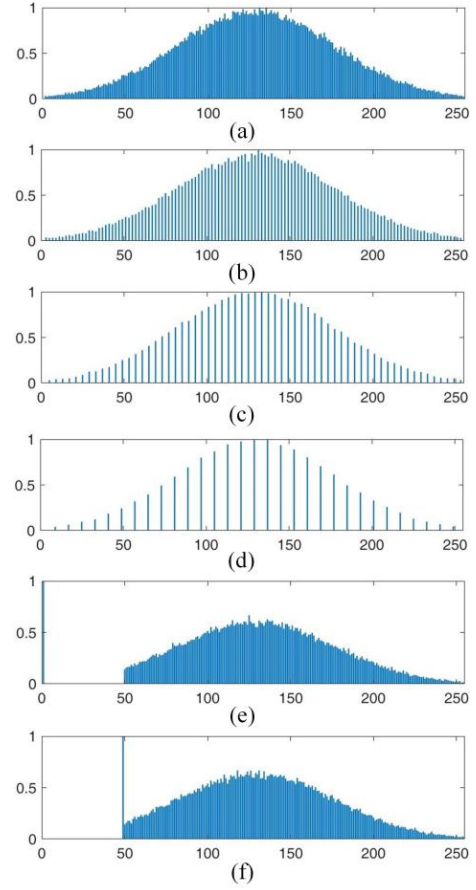

Fig. 1 Gaussian histogram of (a) an image, and its preprocessed versions with (b) $DS_2$, (c) $DS_4$, (d) $DS_8$, (e) $TH_{48}^0$, and (f) $TH_{48}^{48}$.

It is a simple task to demonstrate the effects of applying $DS$ preprocessing to the inputs of a precise computational block (e.g. adder or multiplier) on the Truth-Table (TT) of the resulted PPC block. The Karnough Map (KM) of the third output bit of a 2-bit×3-bit conventional precise multiplier is shown in Fig. 2(a). Fig. 2(b) on the other hand, demonstrates the number and position of the DCs inserted in the KM of the same output bit in a Partially-Precise Multiplier (PPM) whose both inputs are preprocessed with $DS_2$. The KMs of the all output bits of the precise and PPM multipliers can be found in supplementary section III. Generally, the TT of a conventional two input precise computational block consists of $2^{2 \times WL}$ rows with specified values without any DC. However, after applying the $DS_x$ and $DS_{x'}$ preprocessing on its both inputs, the number of DCs in the TT of the resulted PPC block is determined based on the following equation:

Fig. 2 Karnaugh Map of the third output bit in a 2×3 (a) conventional precise multiplier, (b) PPM with $DS_2$ preprocessing on its both inputs, (c) PPM with $TH_5^0$ preprocessing on its 3-bit input, (d) PPM with $TH_5^6$ preprocessing on its 3-bit input.

$$\# \text{ of } DC \text{ rows } = 2^{2 \times WL} \times (1 - (\frac{1}{x} \times \frac{1}{x'})) \quad (1)$$

These DCs highly improve PPC block implementation costs. For example, applying the $DS_2$, $DS_4$ and $DS_8$ on both inputs of a two-input combinational block of any type increases the number of DC rows in the resulted PPC block TT to 75%, 93.75%, and 98.5% respectively. These significant improvements are less or more preserved when implementing the TT in a two-level format or synthesizing it in a multi-level format, depending on the relative position of the DCs in the TT as well as efficiency of the utilized heuristic optimizations in the synthesis tools. As a general rule based on the experimental results, while the number of literals in a two-level implementation of the resulted TT nearly preserves similar savings with respect to number of DCs in TT rows, the area and delay reduction of the multi-level synthesis of the resulted TT provides less improvements as will be explained in the next section.

Also to demonstrate the output accuracy degradation due to applying DS preprocessing on the inputs of a precise computational block, equations 2 and 3 formulate the Probability of Error (PE), Mean Error (ME), and Mean Absolute Error (MAE) of the resulted Partially-Precise Adder (PPA) after applying $DS_x$ on both inputs of a WL-bit precise adder. Equations 4 and 5 also demonstrate the PE, ME, and MAE of a WL-bit PPM when both inputs preprocessed with $DS_x$.

$$PE_{PPA} = 1 - \left(\frac{1}{2^k} \times \frac{1}{2^k}\right) \quad (2)$$

$$ME_{PPA} = MAE_{PPA} = \sum_{i=1}^{2^k} \sum_{j=1}^{2^{WL-k}} \left(\frac{2^{WL}}{2}(2^{WL}-1)(i-1)\right) + (j-1)2^{WL+k-1} = 2^{k-1}(2^{WL-1}-1) + \frac{1}{4} \quad (3)$$

$$PE_{PPM} = 1 - \left(\left(\frac{1}{2^k} \times \frac{1}{2^k}\right) + \left(\frac{2}{2^{WL}} - \frac{2}{2^{k+WL}}\right)\right) \quad (4)$$

$$ME_{PPM} = MAE_{PPM} = 2^{WL+k-1} - 2^{WL-1} - 2^{2WL-2} + 2^{-2} \quad (5)$$

While $k = \log_2 x$ in all above equations.

*2) Thresholding Preprocessing*

The Thresholding preprocessing with two parameters 'x' and 'y' ($TH_x^y$) is another useful preprocessing. The low cost $TH_x^y$ preprocessing maps all input values less than 'x' to the same value 'y'. Figs. 1(e) and (f) demonstrate the normalized Gaussian histogram of the image shown in Fig. 1(a), when preprocessed with $TH_{48}^0$ and $TH_{48}^{48}$ respectively. Increasing the 'x' parameter increases the sparsity that directly affects the resulted PPC precision and cost. The 'y' parameter on the other hand, highly affect the position and not number of DCs in the TT. Therefore, it does not directly affect the input sparsity, while its value indirectly affects the PPC block cost-precision accordingly. The maximum 'x' value as well as the suitable 'y' value in $TH_x^y$ to achieve the best PPC blocks with an acceptable precision for a specific application should be determined based on tolerance of the application against sparsity.

Fig. 2(c), (d) demonstrate the number and position of DCs inserted in the KM of the third output bit in a PPM whose 3-bit input is preprocessed with $TH_5^0$ and $TH_5^6$ preprocessing respectively. The KMs of the all output bits of these PPMs can be found in supplementary section III. Generally, applying the $TH_x^y$ and $TH_{x'}^{y'}$ on both inputs of a WL-bit computational block, reduces the number of its $2^{2 \times WL}$ specified TT rows by introducing many DC rows based on the following equation:

$$\# \text{ of } DC \text{ rows } = 2^{2 \times WL} \times \left(\frac{X}{2^{WL}} \times \frac{X'}{2^{WL}}\right) \quad (6)$$

Equations 7 to 10 also represent the closed forms for the PE, ME, and MAE of a PPA and PPM after applying $TH_x$ preprocessing on both of their inputs.

$$PE_{PPA} = 1 - \left(\frac{x}{2^{WL}} \times \frac{x}{2^{WL}}\right) \quad (7)$$

$$ME_{PPA} = MAE_{PPA} = 2x\sum_{i=x}^{2^{WL}}(i-y) + 2(2^{WL}-x)\sum_{i=x}^{2^{WL}}(i-y) = (1 - \frac{x}{2^{WL}})(2^{WL} + x - 1 - 2M) \quad (8)$$

$$PE_{PPM} = 1 - \left(\left(\frac{x}{2^{WL}} \times \frac{x}{2^{WL}}\right) + 2\left(\frac{2^{WL}-x}{2^{2WL}}\right)\right) \quad (9)$$

$$ME_{PPM} = MAE_{PPM} = 2\left(\frac{2^{WL}-x}{2^{WL}} \times \frac{x}{2^{WL}}\right)\left(\frac{2^{WL}+x-1}{2}\right)\left(\frac{x-1}{2} - M\right) + \left(\frac{2^{WL}-x}{2^{WL}} \times \frac{2^{WL}-x}{2^{WL}}\right)\left(\left(\frac{2^{WL}+x-1}{2}\right)^2 - M^2\right) \quad (10)$$

*3) Down-Sampling and Thresholding Preprocessing Comparison*

Although both of the down-sampling and thresholding preprocessings similarly increase the input signal sparsity to improve the cost of the corresponding PPC blocks, they conceptually differ and also cause different improvements in the achieved PPC blocks. Technically, while the DS increases the input sparsity by means of removing the redundancy between neighbor input values, the TH increases the input sparsity by discarding some irrelevant part of the input data. On the other hand, and from the implementation aspect, while the DS regularly distributes the sparsity among the full input range (as shown in Figs. 1(b)-(d)) with zero implementation costs, the TH irregularly introduces the sparsity only on a specific part of the input range (as indicated in Figs. 1(e) and (f)) by means of a low cost simple circuit. Due to regularity of the output DCs in DS, the conventional synthesis tools and synthesis process can better utilize it to produce more efficient PPC blocks with respect to TH.

III. PPC BLOCK DESIGN FLOW AND IMPLEMENTATION BASED ON NATURALLY EXISTING AND/OR INTENTIONALLY CREATED SPARSITIES

Both of the natural or intentional sparsities can be utilized for development of improved PPC blocks. Fig. 3(a) demonstrates the design-flow of a PPC block of any type as will be explained in the following. The proposed design flow implies the more straightforward development process and usage of customized PPC blocks for an application with respect to imprecise computational blocks. Figs. 3(b) and (c) also demonstrate the implementation process of a PPC block in two-level and multi-level formats respectively as will be explained in detail.

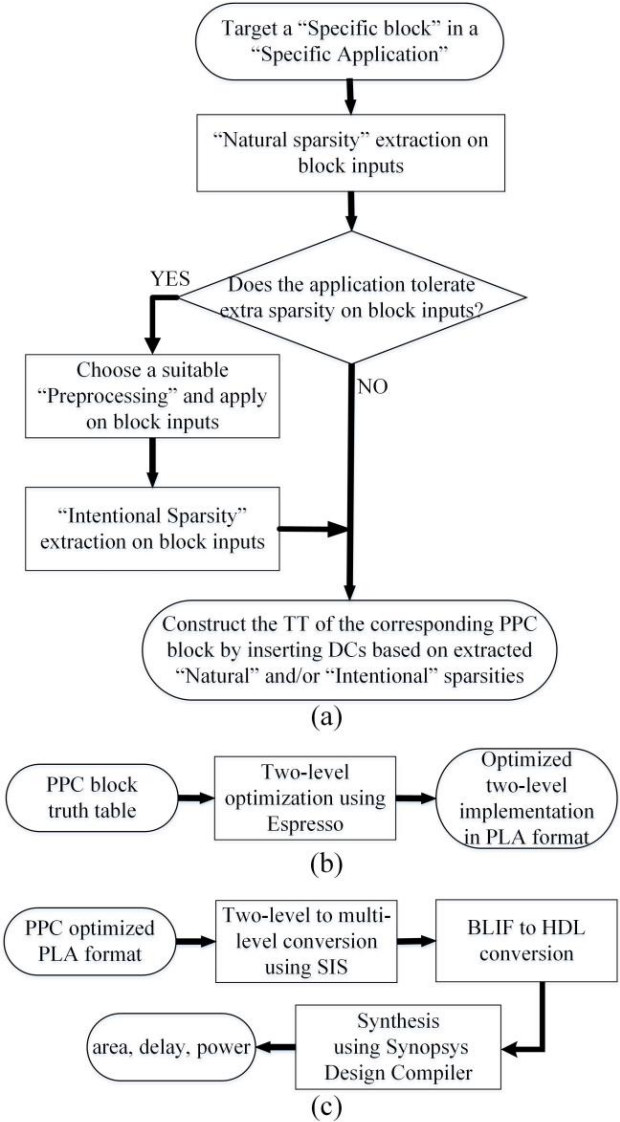

Fig. 3. PPC block (a) Design-flow, (b) two-level implementation process, and (c) multi-level synthesis process.

*A. PPC Block Design Flow*

As indicated in the Fig. 3(a) the design flow of a PPC block launches with a range analysis of the block inputs when it is operating in a specific application, to determine its natural input range sparsity. In other words, the naturally out-of-range input values which essentially do not fed on the inputs of the PPC block in a specific application are first extracted. As the next step, it should be verified whether the specific application can tolerate some intentionally created sparsity with a tolerable accuracy degradation or not. The degradation is caused by intentionally increasing the sparsity on block inputs (in addition to its naturally out-of-range inputs) by applying some type of preprocessing. The suitable type and parameter values of the preprocessing should be determined in a manner to not degrade the overall system quality below an acceptable threshold. As the final step, the extracted "natural" and "intentional" input range sparsities on the block inputs are both utilized to achieve the TT of the simplified PPC block by assigning DCs to the block outputs of the omitted input values. Based on this design flow, as the number of natural or intentional out-of-range input values decreases, the PPC block preserves higher generality while also results in higher implementation cost and performance overheads. On the other hand, increasing the input range sparsity turns the PPC block into a more customized block with improved cost and performance.

*B. PPC Block Two-Level Implementation*

The output TT of the design-flow shown in Fig. 3(a) is considered as the starting point of the implementation process proposed in Fig. 3(b) to extract the two-level hardware implementation of the corresponding PPC block. As shown in Fig. 3(b) a two-level optimization tool such as the Espresso logic optimizer [14], [15], [16] is exploited to minimize the achieved TT of the PPC block in terms of number of literals and produce an optimized SOP model of the desired PPC block in PLA format.

*C. PPC Block Multi-Level Synthesis Process*

The experimental results demonstrate that as the number of DCs increases in the TT, the number of literals of the corresponding PPC block decreases in a linear fashion in its two-level implementation. However, this great achievement cannot be retained when using conventional synthesis process and tools to extract the PPC block area and delay in a multi-level format. The reason is that the conventional synthesis tools exploit a predesigned library of the optimized precise computational blocks to efficiently implement any arithmetic operation. This makes them inherently inadequate for efficient utilization of DCs introduced in TT of a PPC block to optimize its multi-level circuit. Without using this predesigned library, the synthesis tool loses its scalability and provides degraded area/delay efficiencies as the bit-width of the computational block increases due to their internal heuristics and NP-complete algorithms.

To address this drawback two different approaches can be utilized. The first approach is to directly map the preprocessing on the computational block optimized multi-level library structures and omit some parts of the structure. Although this approach results in the most optimized PPCs in terms of their area, delay, and power consumption, it cannot be always utilized due to its two significant disadvantages: 1) it needs human efforts and cannot be applied in an automatic and structured manner and 2) it is not applicable in all preprocessings. For example, while the DS can be directly applied on the optimized structure of a precise array multiplier to achieve the corresponding PPM by means of some human efforts, the TH preprocessing has not this property. The second approach to achieve some efficient PPCs is to modify the conventional synthesis process. The customized synthesis process shown in Fig. 3(c) is proposed and utilized to extract all the implementation results provided in the following sections. It utilizes the DCs in PPC block TT in a more organized manner and thus converts the two-level optimized circuit of a PPC block to a more efficient multi-level format. To extract the PPC block multi-level synthesis results, the two-level optimized PLA file is first fed to the SIS synthesis tool [17]. It accepts the SOP format as input and produces the multi-level optimized circuit of the PPC in a library independent manner in. blif format. A custom parser tool is developed to then transfer the .blif file to VHDL format. It serves as the input of the Synopsys Design Compiler to finally produce the synthesized circuit of the PPC block on TSMC 90nm library. The proposed design flow (Fig. 3(a)) and implementation process (Figs. 3(b) and (c)) can be exploited to develop PPC blocks of different types for any application with various types of sparsity, without any considerations. Although the proposed multi-level synthesis process handles the DCs in a more efficient manner with respect to conventional synthesis process, it suffers from scalability issues as it is expected, due to putting aside the library based synthesis approach. More detailed explanations about the advantages and disadvantages of the proposed synthesis process with respect to the conventional synthesis are provided in the supplementary section II.

## IV. Development of Customized PPC Blocks for Efficient Hardware Implementation of a Gaussian Denoising Filter

The Gaussian Denoising Filter (GDF) is a smoothing operator which is used to blur the image and remove its details and noise [18], [19]. Fig. 4 illustrates the filter coefficients of a discrete GDF filter with the 3x3 window [20]. Fig. 5 also demonstrates the hardware structure of the GDF for denoinsning of a gray-scale image which consists of eight adders with different Word-Lengths (WL). The normalized histograms as well as the WL of all inputs, internal, and output signals of the architecture are also included in the figure. As shown in the figure, all system primary inputs follow a Gaussian histogram due to inherent nature of the input image pixels. The algorithmic weight multiplications are translated onto left-shift operators on inputs of some adders. There are some interesting notes about this figure. The first note is that the 1-bit shift-left operation inserts a $DS_2$-like sparsity on both inputs of the Adder-3 and Adder-4 as shown on their input histograms. The Adder-3, Adder-4, and Adder-6 output histograms also show that these sparsities also propagate deeper to the Adder-7 right input. Similarly, a 2-bit shift-left operation also resembles a $DS_4$-like sparsity as indicated in the histogram of the Adder-8 right input. And finally, it is noteworthy to mention that the algorithmic 1-bit difference between the WLs of the Adder-7 two operands results in a natural-like sparsity on the Adder-7 output (i.e. Adder-8 left input). It is important to note that while the $DS_2$-like algorithmic sparsities are detected and exploited by the synthesis tool to improve the physical properties of the conventionally implemented Gaussian filter, the natural-like sparsity remains unused.

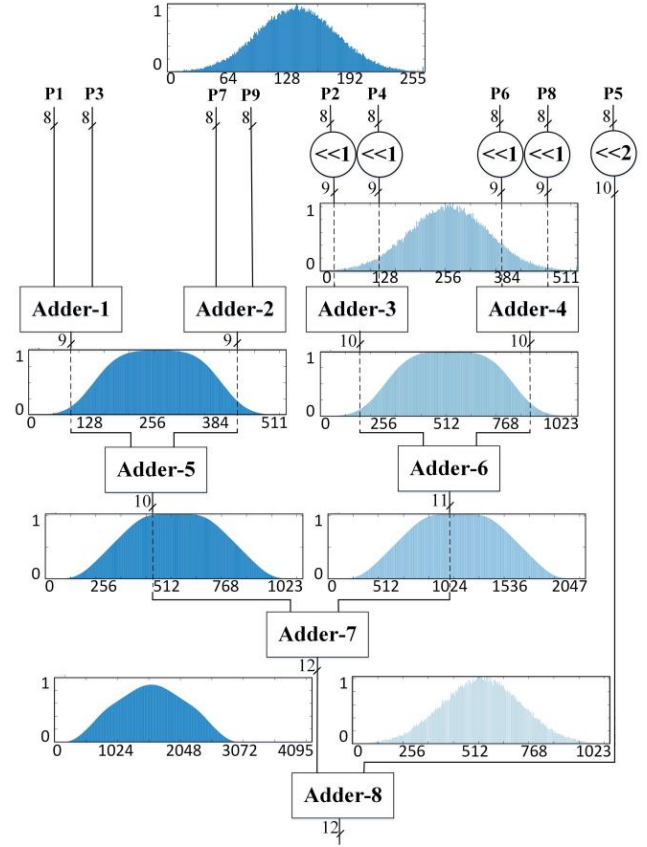

Fig. 4 Window coefficients of the 3x3 Gaussian filter [21]

Fig. 5 Gaussian filter hardware structure, signal Word-Lengths and histograms.

The upper and lower images in Fig. 6(a) illustrate the sample input image and its corresponding filtered output using a conventional precise GDF hardware. Table 1 includes the accuracy as well as physical properties of the conventionally realized and some favorite PPC implementations of the GDF hardware. The first three table columns indicate the number of the row, implementation type (i.e. conventional or PPC), and utilized sparsity type respectively. The next column indicates the output PSNR of the implemented system, while the remaining four columns include the number of literals in two-level implementation, and the area, delay, and power synthesis results respectively. All the implementation results are normalized with respect to the results of the conventional implementation, while the absolute implementation result values are also reported in the supplementary section IV for more convenience. As indicated in table row#1, the conventionally implemented GDF hardware does not utilize any sparsity and provides an ideal infinity PSNR.

Table 1 Cost-accuracy trade-off in conventional and some selected PPC implementations of the Gaussian filter hardware

| Row# | Realization Type | Utilized Sparsity Type(s) | GDF Hardware Accuracy (PSNR) | GDF Hardware Implementation Results | | | |
|---|---|---|---|---|---|---|---|
| | | | | Normalized # of literals | Normalized Area | Normalized Delay | Normalized Power |
| 1 | Conventional | None | Ideal | 1 | 1 | 1 | 1 |
| 2 | Partially-Precise Computing (PPC) | Intentional($DS_2$) | 51 | 0.84 | 0.89 | 1.18 | 0.74 |
| 3 | | Intentional($DS_4$) | 44 | 0.79 | 0.83 | 1.1 | 0.73 |
| 4 | | Intentional($DS_8$) | 37 | 0.73 | 0.72 | 0.95 | 0.66 |
| 5 | | Intentional($DS_{16}$) | 31 | 0.51 | 0.65 | 0.9 | 0.61 |

### A. Gaussian Denoising Filter PPC Implementation Based on Intentionally Created Sparsity

Among different applications which utilize the GDF hardware, there are some cases in which, some extra PSNR degradation is also tolerable at the GDF hardware output with respect to the conventional implementation. This lets to degrade the output by intentionally inserting some sparsity levels on the primary inputs of the GDF hardware by means of a suitable preprocessing. This also helps to replace all existing conventional precise adders with their corresponding PPAs to achieve better costs and performances. The rows 2 to 5 of Table 1 include the accuracy and implementation costs of different Gaussian filter PPC implementations when all the primary inputs of the filter are preprocessed by $DS_x$ for the x values between 2 to 16 respectively. The maximum value of 'x' parameter which resembles maximum tolerable sparsity level and results in the most efficient PPC blocks, should be determined based on the tolerance of an application against inserted sparsity. As indicated in the table, increasing the x parameter both decreases the GDF hardware output PSNR and also improves its number of literals in two-level implementation. However, and due to significant limitations of the existing synthesis tools and processes in efficient utilization of DCs (as explained in Section III.C), this valuable improvement cannot be maintained in all multi-level implementations. The table results show that except for the longer delay of the $DS_2$ and $DS_4$ configurations, the PPC implementations always provide better area, delay, and power consumptions with respect to the conventional implementation. The table accuracy results show that even for the $DS_{16}$ which creates a 93% sparsity on all nine primary inputs of the filter, the resulted output provides 31dB PSNR which is still above 30db and thus categorized as excellent quality [21]. The upper and lower figures in Figs. 6(b) and (c) illustrate the sample preprocessed input image and the corresponding filtered output of the GDF hardware when the filter inputs are preprocessed by $DS_{16}$ and $DS_{32}$ respectively. While Fig. 6(b) shows that the $DS_{16}$ still does not perceptibly degrade the filter input\output images, Fig. 6(c) shows the visible distortions of the input\output images due to applying $DS_{32}$ which degrades the output PSNR to 26dB as a good but not excellent image quality.

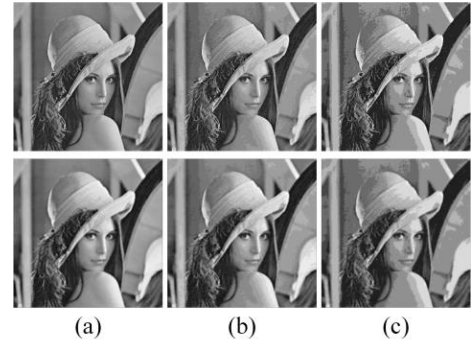

(a) (b) (c)

Fig. 6 Gaussian filter system sample Input/output images for (a) conventional, (b) $DS_{16}$, and (c) $DS_{32}$ implementation.

## V. DEVELOPMENT OF CUSTOMIZED PPC BLOCKS FOR EFFICIENT HARDWARE IMPLEMENTATION OF IMAGE BLENDING APPLICATION

The Image Blending (IB) algorithm is used to mix two input images of the same size with the desired influences. The value of each pixel in the blended output image is a linear combination of the corresponding pixel values of the two input images as follows:

$$P(i.j) = \alpha\, P1(i.j) + (1 - \alpha)\, P2(i.j) \qquad (11)$$

while $P_1$ and $P_2$ are the two input images, P is the blended output, and (i,j) specifies the pixel coordination. α is the blending ratio to determine the influence of each input image on the output. Without losing the system's generality and based on its symmetry, the α value is restricted between 0 and 127 and therefore, 1-α is limited between 128 to 255, considering an 8-bit WL to represent α.

Fig. 7 shows the structure of an IB hardware which blends two gray scale images. The WL as well as sample normalized histogram of each input, internal, and output signal are shown beside each signal in the figure. As the output should be in the range 0 to 255, the 16-bit multiplier output is truncated to 8 bits and then blended using an 8-bit adder as indicated in the figure. As indicated in the figure, the histogram of the image input in both multipliers has a Gaussian shape which covers full input range without any sparsity. On the other hand, the coefficient input histograms of the multiplier-1 and multiplier-2 cover only the first and the last half of the range respectively.

Fig. 8(a) illustrates two sample input images (i.e. Lena and Tulips) and their blended output using a conventional precise hardware when α=0.5. The accuracy and implementation costs of the conventional and some selected PPC implementations of the IB hardware are listed in Table 2. The first three table columns indicate the row number, implementation type (i.e. conventional or PPC), and utilized sparsity type(s) respectively. The fourth column indicates the accuracy analysis of the achieved hardware in terms of its output PSNR. The last four columns also represent the normalized two-level and multi-level implementation results of each implementation. The absolute values of these implementation results are also provided in the supplementary section IV. As indicated in table row#1, the conventionally implemented IB hardware does not utilize any sparsity and provides an ideal infinity PSNR.

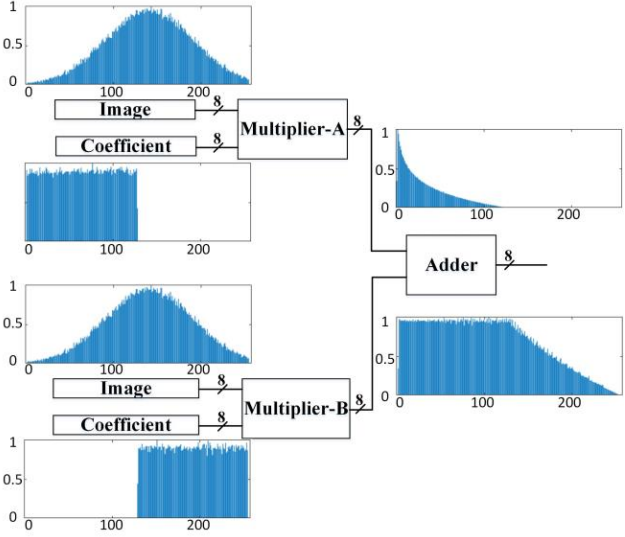

Fig. 7 Image Blending hardware structure, signal Word-Lengths and histograms.

## A. Image Blending PPC Implementation Based on Naturally Existing Sparsity

As indicated in Fig. 7, the coefficient input histograms of both multipliers naturally contain significant sparsity (i.e. exactly half of their range). It can be utilized to develop two different customized PPMs with improved costs in place of conventional precise Multiplier-1 and Multiplier-2. The significant note is that as indicated in the upper adder input histogram, the naturally existing sparsity on the primary inputs of the system also propagated through the multiplier-A to the next level adder and produced significant sparsity levels on its upper input. This lets to replace the precise adder with a customized PPA however, this is not the case here due to its negligible effect on total system cost. Row#2 of Table 2 includes the simulation and synthesis results of the IB hardware, which consists of the two PPM blocks developed based on existing natural sparsity. It has 52% fewer literals, 32% less area, 2% less critical path delay and 49% better power consumption with respect to the conventional implementation. The significant issue is that the PPC blocks developed based on natural sparsity do not degrade the IB system output accuracy with respect to a conventional implementation. Therefore, these savings are achieved without any system PSNR degradation as indicated in the table.

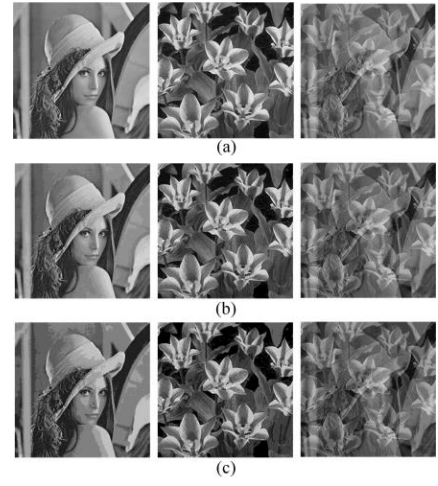

Fig. 8 Image Blending system sample Input/output images for (a) conventional, (b) $DS_{16}$, and (c) $DS_{32}$ implementation.

## B. Image Blending PPC Implementation Based on Intentionally Created Sparsity

Beside the naturally existing sparsity, there are some cases in which, some extra small degradation is also tolerable at the IB hardware output. This helps to develop some more customized and more efficient PPC blocks, this time based on intentional insertion of an extra sparsity on the primary inputs of the IB application by means of some preprocessing, such as $DS_x$ in this case. Rows #3 to #7 of Table 2 present the accuracy and implementation results of the IB hardware (Fig. 7) composed of two PPMs and a precise adder. This time, the PPMs are developed based on intentional sparsity inserted by applying $DS_2$, $DS_4$, $DS_8$, $DS_{16}$, and $DS_{32}$ preprocessing on both image and coefficient inputs of the two multipliers. It is important to consider that the table results demonstrate the superiority of the PPC versions in terms of their physical properties with respect to conventional implementation, except for the delay in $DS_2$, $DS_4$, and $DS_8$. The table accuracy analysis results show that the IB hardware output PSNR degrades more as the preprocessing parameter increases. However, even the $DS_{16}$ which needs 79% less number of literals, 62% better area, 9% less delay, and 73% better power consumption with respect to the conventional implementation, provides an excellent 30dB PSNR. Figs. 8(b) and (c) illustrate the two sample input images and their corresponding outputs when they are preprocessed by $DS_{16}$ and $DS_{32}$ respectively. Fig. 8(b) shows that the $DS_{16}$ does not degrade the input images and thus the system output in a sensible way. However, Fig. 8(c) demonstrates that the $DS_{32}$ causes visible distortions on the input images and thus degrades the output PSNR to 23dB which is below the excellent quality threshold.

Table 2 Cost-accuracy trade-off in conventional and some selected PPC implementations of the IB hardware.

| Row# | Realization paradigm | Utilized Sparsity Type(s) | IB Hardware Accuracy (PSNR) | IB Hardware Implementation Results | | | |
|---|---|---|---|---|---|---|---|
| | | | | # of literals | Area | Delay | Power |
| 1 | Conventional | None | Ideal | 1.000 | 1.00 | 1.00 | 1 |
| 2 | Partially-Precise Computing (PPC) | Natural | Ideal | 0.486 | 0.68 | 0.98 | 0.51 |
| 3 | | Intentional ($DS_2$) | 49 | 0.298 | 0.75 | 1.13 | 0.55 |
| 4 | | Intentional ($DS_4$) | 42 | 0.084 | 0.67 | 1.11 | 0.47 |
| 5 | | Intentional ($DS_8$) | 39 | 0.031 | 0.53 | 1.11 | 0.40 |
| 6 | | Intentional ($DS_{16}$) | 30 | 0.021 | 0.38 | 0.91 | 0.27 |
| 7 | | Intentional ($DS_{32}$) | 23 | 0.019 | 0.15 | 0.46 | 0.09 |
| 8 | | Natural & Intentional ($DS_2$) | 49 | 0.154 | 0.60 | 0.93 | 0.44 |
| 9 | | Natural & Intentional ($DS_4$) | 42 | 0.052 | 0.46 | 0.95 | 0.36 |
| 10 | | Natural & Intentional ($DS_8$) | 39 | 0.025 | 0.38 | 0.93 | 0.29 |
| 11 | | Natural & Intentional ($DS_{16}$) | 30 | 0.020 | 0.27 | 0.75 | 0.17 |

*C. Image Blending PPC Implementation Based on both Natural and Intentional Sparsity*

Table 2 rows #2 to #7 show that considering either of the natural or intentional sparsities for development of customized PPC blocks considerably improves the system's physical properties. However, simultaneously considering both the natural and intentional sparsities is a better idea which more intensifies the improvements. The last four rows of Table 2 include the simulation and synthesis results of the IB hardware when it is composed of PPC multipliers developed based on the union of natural sparsity as well as the intentional sparsity created by down-sampling with different rates. Comparison of the results in rows #3 to #6 with their counterparts in rows #8 to #11 clearly demonstrates that for any DS parameter value, utilizing the natural sparsity does not degrade the system's accuracy while highly improves its implementation costs. For example, rows #5 and #10 of the table demonstrate that utilizing the natural sparsity along with $DS_8$ does not degrade the system accuracy with respect to only $DS_8$, while more improves the number of literals, area, delay, and power consumption by about 21%, 29%, 16%, and 27% respectively.

## VI. Development of Customized PPC Blocks for Efficient Hardware Implementation of a Face Recognition Neural Network

The hardware structure of the implemented three-layer Face Recognition Neural Network (FRNN) is demonstrated in Fig. 9. It consists of 960 inputs based on the size of the dataset images (32 pixels by 30 pixels), 40 hidden neurons, and 7 output neurons (4 for ID, 2 for direction, and 1 for sunglass detection of the input face image). A sample input image of the network is also shown in Fig. 11 (a). Each neuron consists of a hardware MAC and a sigmoid transfer function. Fig. 10 demonstrates the internal structure of the utilized MAC, excluding the transfer function. The WLs and normalized histograms of the input and internal signals of the MAC are also illustrated in the figure beside each signal. The histogram shown for the multiplier image input is the union of the histograms of all the dataset images. The histogram of the adder upper input (or the multiplier output) and the adder feedback input are also demonstrated in the figure. It should be noted that both the utilized network and its dataset are developed and released years ago [22] without any contribution from the existing work.

Table 3 also demonstrates the accuracy metrics and implementation results of the conventional as well as some selected PPC versions of the FRNN. The first three table columns indicate the row number, implementation type (i.e. conventional or PPC), and utilized sparsity type(s). The next three columns indicate the accuracy metrics of the resulted FRNN in terms of its Correct Classification Rate (CCR), Training Epoch (TE) and output Mean-Squared Error (MSE). The last four table columns represent the normalized values of the number of literals in two-level implementation, and the area, delay, and power consumption synthesis result respectively. The implementation results of these three columns also can be found in the supplementary section IV. As shown in the first table row, the conventionally implemented FRNN hardware does not utilize any sparsity, provides a correct classification rate of 89%, has an output MSE of 0.026, and needs 170 training epochs.

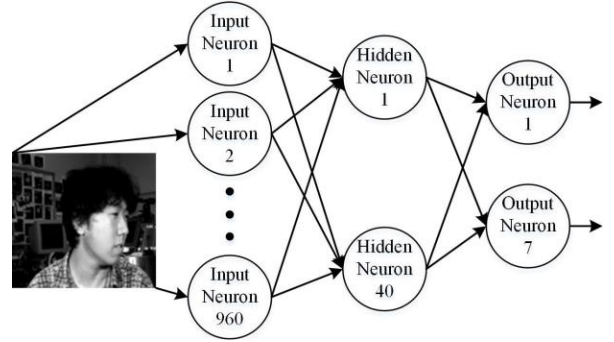

Fig. 9 Whole structure of Face Recognition Neural Network

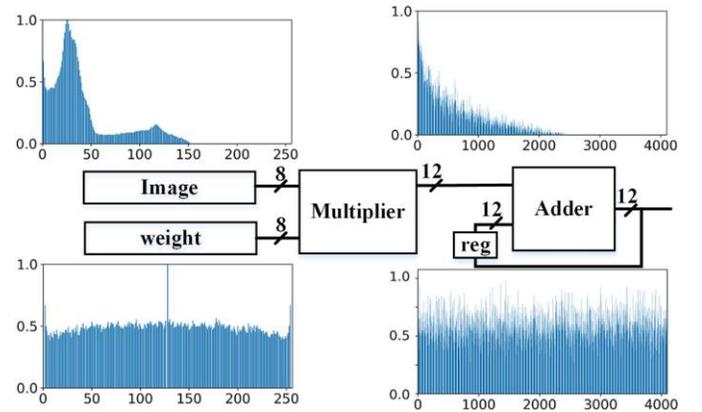

Fig. 10 Face Recognition Neural Network neuron structure, signal Word-Lengths and histograms.

*A. Face Recognition PPC Implementation Based on Naturally Existing Sparsity*

The two input histograms of the MAC multiplier (shown in Fig. 10) demonstrate that while the multiplier weight input histogram covers the entire range, the pixel values between 160 and 255 do not appear on the multiplier image input. This natural sparsity helps to replace the precise multiplier with a customized PPM without any accuracy degradation. The adder upper input histogram also shows that this natural sparsity has also been transferred to the adder input through the multiplier. This guides to replace the next level precise adder with a customized PPA. However, all different FRNN PPC versions implemented in this section are all constructed by means of a precise adder. The second row of Table 3 includes the accuracy simulation and implementation results of the FRNN, while its hidden layer neurons are constructed using PPMs developed based on the existing natural image sparsity. The results show that it provides 38% fewer literals with respect to the conventionally implemented network without any accuracy degradation. However, this valuable improvement cannot be maintained in multi-level implementation (as explained before) and the synthesized circuit needs 19% more area, 25% longer critical path, and 3% worse power consumption.

Table 3 Cost-accuracy trade-off in conventional and some selected PPC implementations of the FRNN hardware

| Row# | Realization paradigm | Utilized Sparsity Type(s) | FRNN Accuracy Simulation Results | | | Single Neuron (MAC) Implementation Results | | | |
|---|---|---|---|---|---|---|---|---|---|
| | | | CCR | TE | MSE | # of literals | Area | Delay | Power |
| 1 | Conventional | None | 89 | 170 | 0.02 | 1.000 | 1.000 | 1.000 | 1 |
| 2 | Partially Precise Computing (PPC) | Natural | 89 | 170 | 0.02 | 0.625 | 1.198 | 1.258 | 1.03 |
| 3 | | Intentional ($TH_{48}^{48}$) | 87 | 240 | 0.03 | 0.882 | 1.496 | 1.459 | 1.22 |
| 4 | | Intentional ($DS_{16}$) | 87 | 330 | 0.04 | 0.019 | 0.431 | 0.983 | 0.35 |
| 5 | | Intentional ($DS_{32}$) | 82 | 600 | 0.04 | 0.017 | 0.241 | 0.580 | 0.18 |
| 6 | | Natural & Intentional ($DS_{16}$) | 87 | 330 | 0.04 | 0.018 | 0.383 | 0.812 | 0.27 |
| 7 | | Natural & Intentional ($DS_{32}$) | 82 | 600 | 0.04 | 0.017 | 0.221 | 0.543 | 0.16 |
| 8 | | Natural & Intentional ($TH_{48}^{48} + DS_{16}$) | 82 | 250 | 0.04 | 0.018 | 0.410 | 0.885 | 0.30 |
| 9 | | Natural & Intentional ($TH_{48}^{48} + DS_{32}$) | 85 | 250 | 0.04 | 0.017 | 0.220 | 0.524 | 0.16 |

*B. Face Recognition PPC Implementation Based on Thresholding Sparsity*

Beside the naturally existing input sparsity, it is also possible to more increase the FRNN input sparsity by intentionally discarding some irrelevant information of the input images. As an instance, the dark background of the face images does not contribute to the recognition of the face and therefore is a good removal candidate by means of thresholding preprocessing. To efficiently determine the maximum 'x' parameter value in $TH_x^y$ preprocessing (to achieve the maximum sparsity with still tolerable damage on the face image), Fig. 12 (a) demonstrates the CCR and MSE of the FRNN for different threshold (i.e. 'x') values. As indicated in this figure, choosing a threshold value of about 48 (indicated by the vertical dashed line on the figure) does not significantly change the FRNN MSE and CCR, while inserts about 19% (i.e. 48/256) sparsity on the multiplier image input. As the value of the 'y' parameter does not affect the sparsity of the signal, its value is chosen as the least existing pixel value in the preprocessed images i.e. 48.

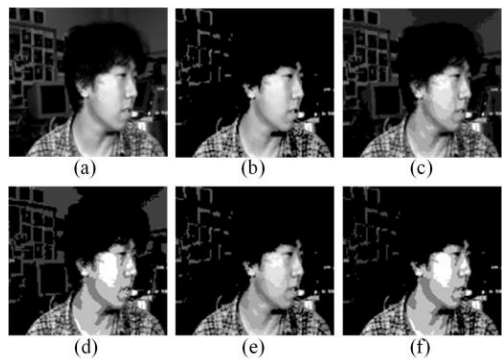

Fig. 11 Face recognition neural network hardware sample input image for (a) precise (b), $TH_{48}^{48}$, (c), $DS_{16}$, (d), $DS_{32}$, (e), mixed natural, $DS_{16}$ and $TH_{48}^{48}$, and (f), mixed natural, $DS_{32}$ and $TH_{48}^{48}$ implementations.

Therefore, the irrelevant background of the images can be omitted by mapping all dark pixels of the image with values

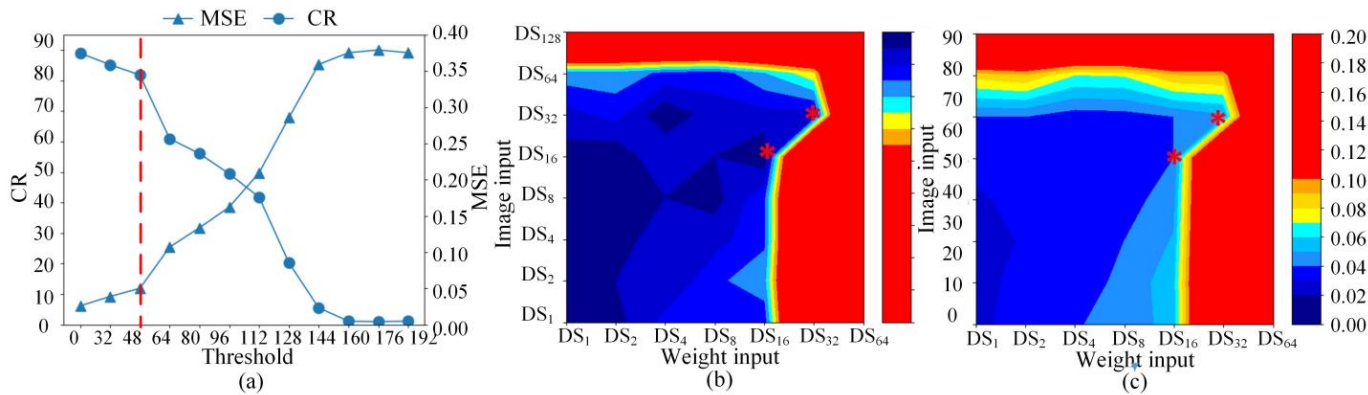

Fig. 12 (a) CCR and MSE of the FRNN when applying $TH_x^0$ preprocessing with different threshold values (i.e. 'x') on image input of the utilized PPC multiplier. (b), (c) CCR and MSE of the FRNN when applying DSx preprocessing with different 'x' parameter values on both image input (vertical axis) and weight input (horizontal axis) of the PPC multiplier respectively.

less than 48 to value 48 by means of $TH_{48}^{48}$ preprocessing. This negligibly affects the face image (as shown in Fig. 11 (b)), as the main significant object which affects the FRNN functionality, while highly increases the multiplier image input sparsity. The third row in Table 3 includes the implementation results of the FRNN composed of the PPM constructed based on intentional sparsity created by means of $TH_{48}^{48}$ preprocessing. It provides only 2% less CCR with respect to the conventional implementation, while provides 12% fewer literals in a two-level implementation. However, the introduced DCs cannot be efficiently utilized in a multi-level synthesis process, which results in 49% more area, 45% longer delay, and 4% more power consumption with respect to conventional implementation.

*C. Face Recognition PPC Implementation Based on Down-Sampling Sparsity*

Some further experiments also show that the FRNN can still efficiently recognize the faces even after removing some redundant pixels from the input images by means of $DS_x$ preprocessing. To achieve the PPC blocks with the least possible cost and good enough precision for an specific application, the maximum DS parameter should be chosen based on the tolerance of that specific application against input sparsity as explained before. To determine the maximum 'x' parameter value in $DS_x$ preprocessing in FRNN, Figs. 12(b) and (c) illustrate the CCR and MSE of the FRNN when similar down-sampling rates (i.e. 'x' parameter values) are simultaneously applied to the image and weight inputs of the multiplier respectively. The point relates to ($DS_1$, $DS_1$) corresponds to the conventional FRNN implementation using precise computational blocks. The red regions on both maps correspond to the configurations in which the FRNN cannot be trained at all, due to unacceptable and very high down-sampling of either of the inputs. The figures demonstrate that increasing the down-sampling rate for each one of the multiplier inputs degrades the FRNN CCR and MSE accuracy metrics. The figures also clearly demonstrate that the image input of the multiplier can be preprocessed with respectively higher down-sampling rates than the weight input. The red stars on both figures indicate two selected PPC configurations with the highest possible down-sampling rates and still acceptable accuracies. These two points correspond to applying $DS_{16}$ and $DS_{32}$ on both multiplier inputs respectively.

Rows #4 and #5 of Table 3 include the accuracy and implementation results of the FRNN hardware when applying $DS_{16}$ and $DS_{32}$ on both multiplier inputs. Figs. 11(c) and (d) demonstrate the sample image inputs after applying these two preprocessing respectively. The table results show that removing the redundant data from the image and weight inputs by means of $DS_{16}$ preprocessing provides 2% worse CCR with respect to conventional implementation. While significantly decreases the number of possible input values by a factor of 1/256 and thus requiring 98.1% fewer literals. The synthesis results show that it also needs 57% less area, provides 2% better delay and 65% less power consumption with respect to the conventional implementation. Similarly, although applying the $DS_{32}$ to the image and weight inputs degrades the CCR about 8%, it considerably drops the number of possible input values by a factor of 1/1024 and requires 98.3% fewer literals with respect to conventional implementation in a two-level format. It also provides 76%, 42%, and 82% better area, delay and power respectively. As indicated in rows #2 to #5 of Table 3, utilizing each one of the natural and intentional sparsities can individually improve the number of FRNN literals in two-level implementation. However, these significant savings cannot be preserved in multi-level synthesis process in natural and thresholding sparsities as indicated in rows #2 and #3 of the table. On the other hand, rows #4 and #5 of the table demonstrate that the PPC blocks developed based on DS preprocessing provide great advantages even in multi-level implementation due to regularity of the sparsity and DCs introduced by DS preprocessing as explained before.

*D. Face Recognition PPC Implementation Based on Naturally existing, Thresholding, and Down-Sampling Sparsity*

To utilize more possible sparsity and thus extract more efficient PPC blocks for the FRNN, rows #6 and #7 of the table demonstrate the accuracy and implementation results of different combinations of the mixed natural and intentional DS preprocessing. As indicated in the table, adding the natural sparsity to an intentional sparsity does not affect its accuracy as expected, while improving its implementation costs. For example, a comparison between results in rows #5 and #7 clearly shows that the mixed natural and $DS_{32}$ sparsities provides the same accuracy as the $DS_{32}$ while improves its number of literals, area, delay, and power by about 1%, 9%, 5%, and 11% respectively. The same Figs. 11(c) and (d) can also be referred, to illustrate the corresponding preprocessed input image samples for PPC configurations addressed in rows #6 and #7 of the table. Finally, the last two table rows include the accuracy results and implementation costs of the FRNN composed of the PPC blocks which are developed based on mixed natural, thresholding, and down-sampling preprocessings. Figs. 11(e) and (f) illustrate the corresponding preprocessed input image samples for these two configurations. The last table row indicated that a mixed natural, $TH_{48}^{48}$, and $DS_{32}$ provides 4% less CCR with respect to the conventional implementation, while results in a tremendous 98% reduction in the number of literals, 78% less area, 48% less delay, and 84% better power consumption.

VII. CONCLUSION

Based on an inspiration from the biological information reduction hypothesis, a novel computational paradigm named as bioinspired partially-precise computing is introduced in this paper. A bioinspired partially-precise computational block is specifically developed for a specific application to provide the correct result only for a limited set of task-relevant inputs, based on requirements of the specific application. In this paper, the first instances of PPC adders/multipliers are introduced for the image blending and face recognition neural network embedded applications. The experimental results for image blending show that the proposed hardware provides 98% less number of literals with respect to a conventional precise system, while the output PSNR of application remains over 30 dB. Also the proposed PPC blocks for the face recognition neural network provide 98% less literals with respect to a conventional precise system with only 4% correct classification rate reduction.

# Supplementary

I. CUSTOMIZED VISUAL SYSTEM OF THE CREATURES AND ITS SIGNIFICANT IMPACT ON THEIR DAILY LIFE

The color vision of the human and animals is an interesting instance of dimensionality reduction. Fig. 1 illustrates a comparative study between visual system specifications and output of the human vs some animals. Fig. 1(a) demonstrates the spectral sensitivity curves of the human eye as well as five different snapshots of the human eyesight when looking at different natural scenes. On the other hand, five different columns of the Fig. 1(b) show the spectral sensitivity curves of the five animals (i.e. bee, pigeon, mantis shrimp, snake, and shark from left to right respectively). To better demonstrate the effects of each animal's spectral sensitivity curve on its visual system output and compare it against a human, the snapshot at the bottom of each column includes the animal eye output when looking at the similar scene indicated above in Fig. 1(a).

A natural light beam inherently consists of a very wide range of frequencies and can be described in terms of its power spectrum across all frequencies by means of a spectrophotometer [12]. As shown in the sensitivity diagram of Fig. 1(a), the human retina has only three kinds of color sensors called cones, to only detect red, green, and blue color wavelengths. Consequently, human color perception is customized to only cover a three-dimensional representation of the original light with an infinite number of dimensions represented by means of many frequencies [8]. The preprocessed information is then transformed to its main processing engine (i.e. the brain) for different types of further in-depth conceptual processing.

The spectral sensitivity curves of some animals shown in Fig. 1(b) indicate fundamental hardware differences between their visual systems, which also significantly affects their visual system output as shown in the corresponding sample snapshots. The reason is that, the eyes and in a more general way the sensory equipment as well as processing engine of each creature are all customized in a manner to efficiently represent and respond to only a relatively small set of behaviorally meaningful vital inputs [8], within the around rich and varying natural world. This simplifies the brain's computation and limits its operation only to a selected significant set of inputs. Some detailed justifications about vital importance of these custom spectral sensitivity curves in each animal's survival is provided as the following.

*A. Bee*

As indicated by the first column of Fig. 1b, the spectral sensitivity curve as well as the sample bee color vision snapshot reveal that the bee color vision is very different from that of a human [1]. While the human eyes have three types of cones sensitive to red, green and blue colors, the honeybees have receptors for only green, blue and ultraviolet. This causes a shift in the wavelength range they can see [2]. This biological phenomenon can be justified based on vital requirements of the bees: the flowers that try to attract bees and other pollinators often have details that can only be seen with UV-vision [3]. Moreover, the uniformly spaced receptors of the honeybees are also optimal for discriminating flower colors [4].

*B. Bird*

The spectral sensitivity curve in the second column of Fig. 1b demonstrates that the bird color vision consists of four evenly spaced relatively narrow response curves with less redundancy or overlap in comparison with that of a human [2]. Moreover, As the leaf spectra are more variable above the 555 nm, the bird color vision also includes some long-wavelength cones with maximum sensitivity of around 600 nm [4] which results in a more clear vision of the leaves as illustrated in the presented snapshot from the bird color vision. Furthermore, the bird color vision includes a UV channel which facilitates navigation through complex natural leafy environments, as well as the localization of particular leaf surfaces for various tasks, including prey searching, oviposition, and refuge seeking [5].

*C. Mantis Shrimp*

The third column in Fig. 1b demonstrates the Mantis-shrimp color vision spectral sensitivity and sample snapshot. The provided spectral sensitivity diagram shows that its color vision includes the highest number of photoreceptor types ever known among other animals [6]. There are some species with about 12 to 16 different receptors which enables them to discriminate many colors with near wavelengths. Also the Mantis-shrimp does not have the ability to discriminate between closely positioned wavelengths, its color vision system enables it to make quick and reliable determinations of color with a very low processing delay. Due to the Mantis-shrimp live a rapid-fire lifestyle, this simple and fast yet temporally efficient color recognition ability is critical for its survival [6].

*D. Snake*

Fig. 1b fourth column demonstrates the spectral sensitivity diagram as well as a sample real world snapshot of a snake. The night activity is a vital requirement of a snake due to the nocturnal behavior of its primary preys. Therefore, the snake color vision is augmented with ultraviolet-sensitive photoreceptors as indicated in the spectral sensitivity diagram, which perfectly customize it for foraging and ambushing in a dim light environments [7]. The provided snapshot clearly demonstrates how the thermal sensitive like color vision of a snake filters the undesired information, to simplify the distinction process and even size estimation of a probably live prey in its brain as its most important vital requirement [7].

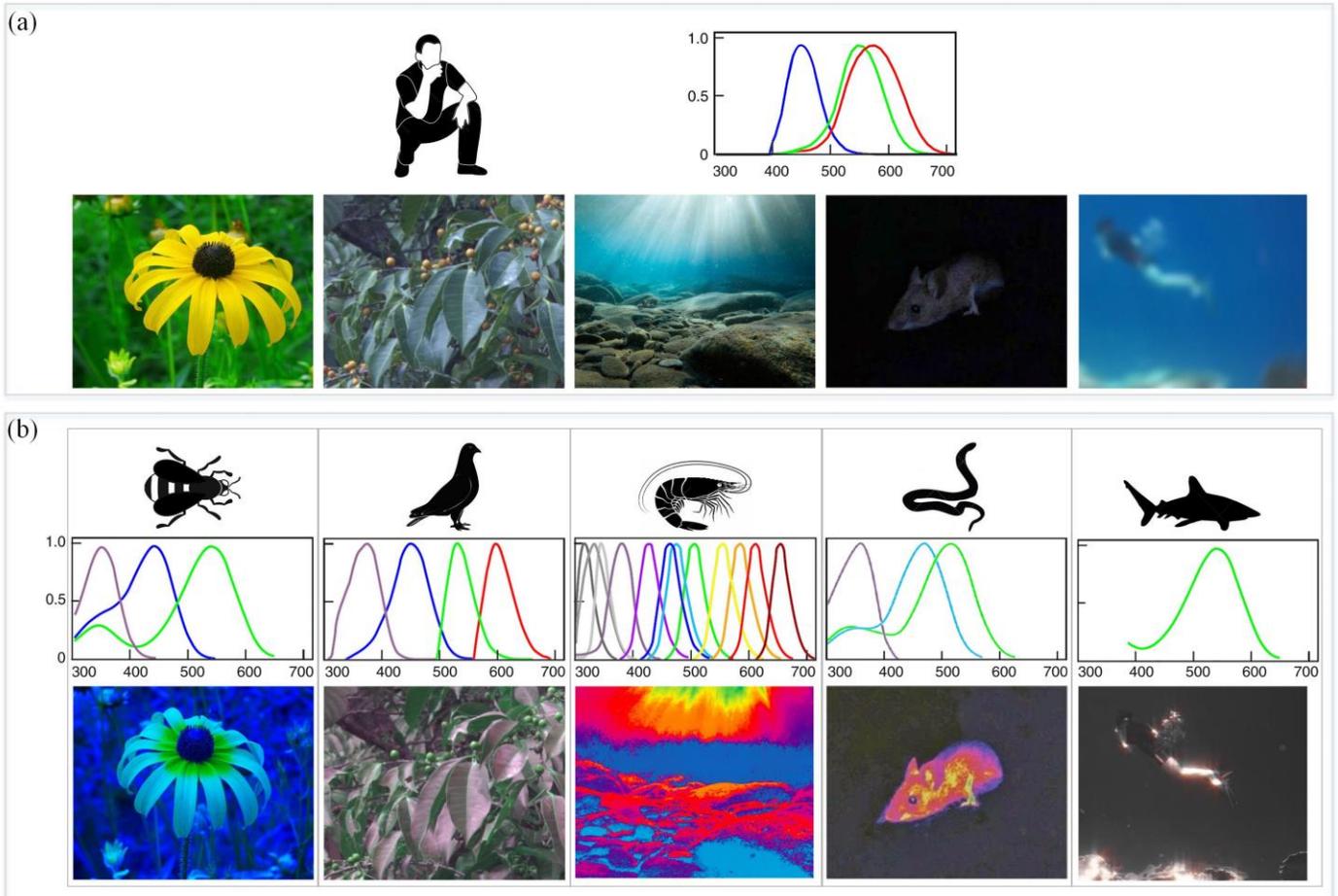

Fig. 1. Visual system spectral sensitivity curve and sample snapshots [9] of the (a) Human, and (b) five different animals including Bee, Bird, Mantis-shrimp, Snake, and Shark from left to right.

*E. Shark*

And finally the last Fig. 1b column illustrates the color vision spectral sensitivity curve and sample scene snapshot of a shark. A shark inhabit a limited water depths of the seas and oceans while it should recognize the objects through long distance underwater [2]. As it should survive in such a low contrast visual environment, its retina is customized to provide very high contrast sensitivities which is vital for object detection and discrimination underwater. While a complex color vision potentially degrades spatial acuity, the shark retina is consists of only one type of cones to be able to provide high resolution achromatic vision signals [8].

## II. ADVANTAGES AND DISADVANTAGES OF THE PROPOSED SYNTHESIS PROCESS

To provide a better understanding about the pros\cons of the conventional and proposed synthesis processes, Table 1 demonstrates the area/delay of some precise signed and unsigned 8-bit×8-bit multipliers, synthesized using both conventional and the proposed processes. To demonstrate the capabilities of the conventional and proposed synthesis processes in handling DCs in the outputs of the multipliers, the physical properties of multipliers are reported in three cases when the number of required multiplier output bits are 16, 12, and 8 bits as indicated in the second table column. The zero, four and eight least significant bits of the 8×8 multiplier output can be considered as DC when the output WL is considered as 16, 12 and 8 respectively. Also as direct applying the proposed synthesis process to an 8×8 multiplier results in very high area/delay due to its scalability issues, the structure shown in Fig. 2 is exploited only for the proposed approach to address the scalability problem. In this scheme, the 8×8 multiplier is composed of four 4×4 multipliers which can be efficiently handled using the proposed synthesis process. The resulted partial products are then directly added to achieve the final result.

Table 1: Area/delay synthesis results of 8-bit × 8-bit signed and unsigned precise multipliers with different output WLs, using conventional and proposed synthesis processes.

| Multiplication Operand Type | Output WL | Conventional Synthesis Process | | Proposed Synthesis Process | |
|---|---|---|---|---|---|
| | | Area (GE) | Delay (ns) | Area (GE) | Delay (ns) |
| Unsigned | 16 | 1143 | 2.17 | 1855 | 3.38 |
| | 12 | 1131 | 2.07 | 1621 | 3.04 |
| | 8 | 1130 | 2.18 | 987 | 2.44 |
| Signed | 16 | 1216 | 2.13 | 1743 | 3.37 |
| | 12 | 1202 | 2.08 | 1651 | 3.02 |
| | 8 | 1201 | 2.14 | 1039 | 2.55 |

The table results demonstrate that in both signed and unsigned multipliers, when there is no DC on the multiplier output (i.e. the output WL is 16), the proposed synthesis process results in meaningfully worse area/delay with respect to conventionally synthesized multiplier due to the overheads of the structure shown in Fig. 2 with respect to a conventional 8×8 multiplier. The significant note about the table results is that increasing the number of DCs on the multiplier outputs to 4 and 8 (i.e. decreasing the output WL to 12 and 8 respectively) does not meaningfully affect the area/delay of the multiplier in a conventional synthesis process due to library based approach. However, the proposed synthesis process can efficiently utilize the output DCs and the area/delay of the multiplier decreases as the number of DCs on its output increases more. The last interesting note about the table results is that while in a conventional synthesis process the signed multipliers have slightly more area with respect to the unsigned multipliers with the same output WL due to its optimum pre-designed library components, the proposed synthesis process does not result is a meaningful difference between the area/delay of the signed and unsigned multipliers due to its TT based approach.

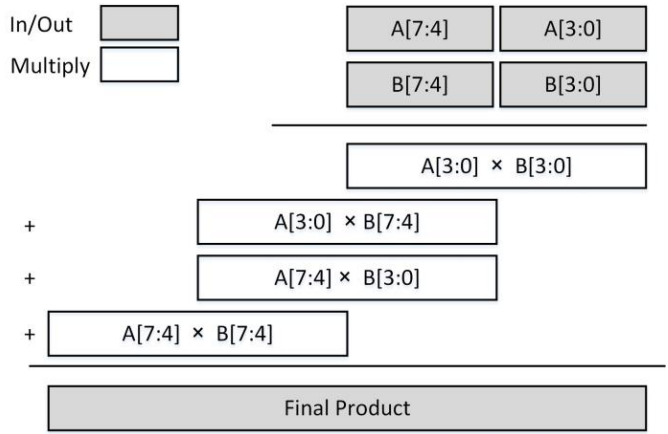

Fig. 2 Construction of an 8×8 multiplier by utilizing four 4×4 multipliers to address the scalability issues of the proposed multi-level synthesis process.

In a similar manner, as direct application of the proposed synthesis process on different 8-bit to 12-bit adders in the Gausian filter, results in very high area/delay due to scalability issues of the proposed synthesis process, all utilized adders are constructed by cascading smaller 4-bit segments which can be efficiently handled using the proposed synthesis process to overcome this drawback. The structure of the utilized 12-bit adders in the Gaussian filter is shown in Fig. 3 as an instance. It should be noted that the two other applications (i.e. image blending hardware and face-recognition neural network) utilize the conventional precise adder instead of PPAs due to its very low impact on physical properties of the whole design with respect to the existing multipliers as indicated in the paper.

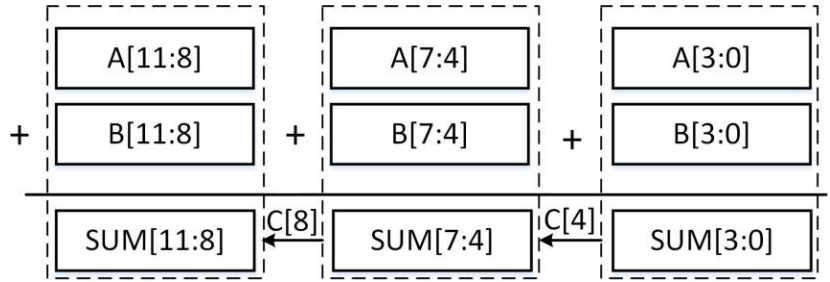

Fig. 3 Construction of a 12-bit Adder by cascading 4-bit Adders to address the scalability issues of the proposed multi-level synthesis process.

III. KARNOUGH MAPS OF THE OUTPUT BITS OF 2X3 PRECISE AND BPM MULTIPLIERS

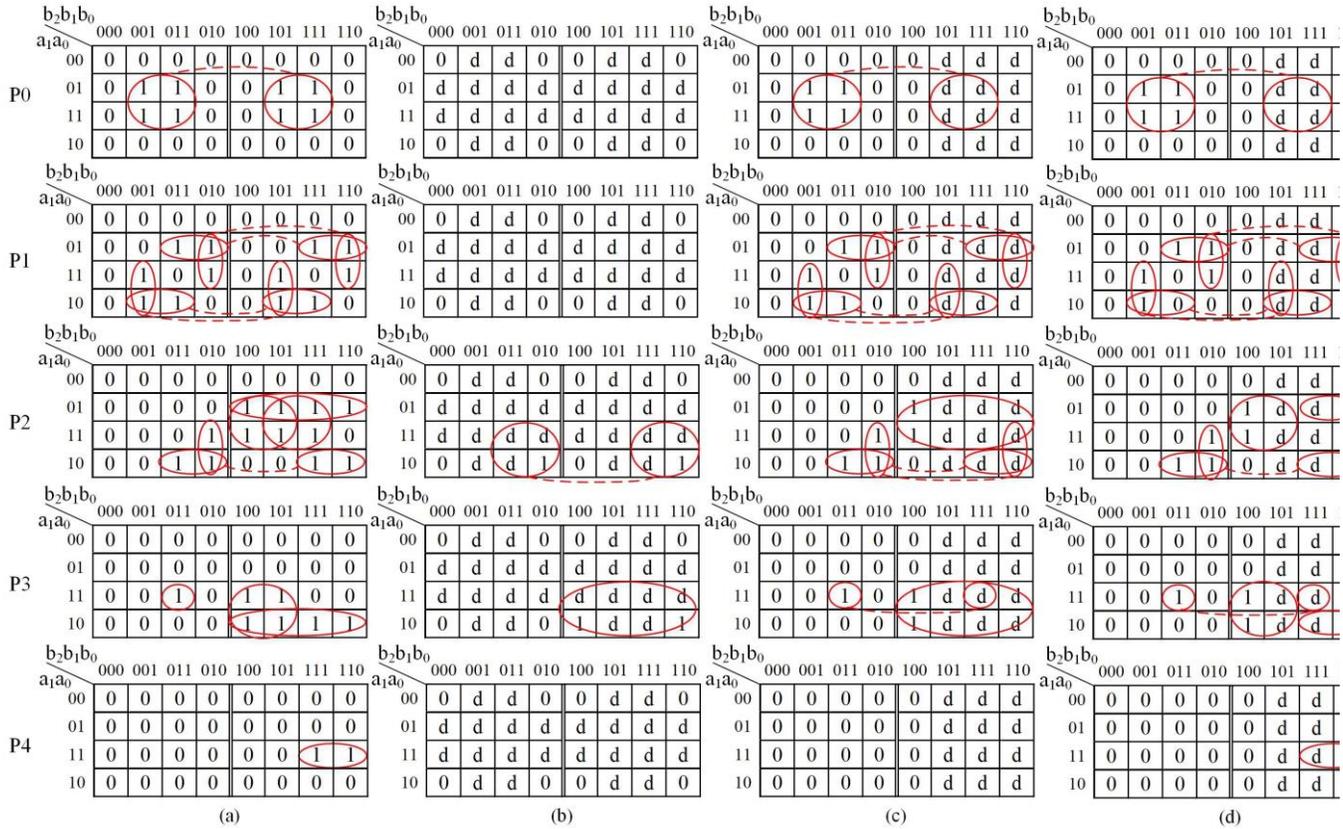

Fig. 4 Karnough maps of the all output bits of a 2x3 (a) precise multiplier, (b) BPM with $DS_2$ preprocessing on its both inputs, (c) BPM with $TH_5^0$ preprocessing on its 3-bit input, (d) BPM with $TH_5^6$ preprocessing on its 3-bit input

IV. Absolute Implementation Results of the Image Blending and Face Recognition Neural Network Precise and BPC Implementations

Table 2 Cost-accuracy trade-off in conventional and some selected BPC implementations of the Gaussian Filter hardware (absolute implementation result).

| Row# | Realization Type | Utilized Sparsity Type(s) | GDF Hardware Accuracy (PSNR) | GDF Hardware Implementation Results | | | |
|---|---|---|---|---|---|---|---|
| | | | | Two-Level (# of literals) | Area (# of GE*) | Delay (ns**) | Power (uw***) |
| 1 | Conventional | None | Ideal | 7658 | 1145 | 3.87 | 102 |
| 3 | Bioinspired Partially Precise Computing (BPC) | Intentional($DS_2$) | 51 | 6468 | 1018 | 4.58 | 75 |
| 4 | | Intentional($DS_4$) | 44 | 6048 | 955 | 4.26 | 75 |
| 5 | | Intentional($DS_8$) | 37 | 5581 | 827 | 3.68 | 67 |
| 6 | | Intentional($DS_{16}$) | 31 | 3892 | 744 | 3.48 | 62 |

*Number of Gate Equivalents, **Nanoseconds, ***Micro Watt

Table 3 Cost-accuracy trade-off in conventional and some selected BPC implementations of the IB hardware (absolute implementation result).

| Row# | Realization Type | Utilized Sparsity Type(s) | IB Hardware Accuracy (PSNR) | IB Hardware Implementation Results | | | |
|---|---|---|---|---|---|---|---|
| | | | | Two-Level (# of literals) | Area (# of GE*) | Delay (ns**) | Power (uw***) |
| 1 | Conventional | None | Ideal | 541792 | 2404 | 3.01 | 461 |
| 2 | Bioinspired Partially Precise Computing (BPC) | Natural | Ideal | 263189 | 1641 | 2.96 | 235 |
| 3 | | Intentional($DS_2$) | 49 | 161266 | 1814 | 3.41 | 253 |
| 4 | | Intentional($DS_4$) | 42 | 45460 | 1612 | 3.35 | 216 |
| 5 | | Intentional($DS_8$) | 39 | 17036 | 1284 | 3.36 | 185 |
| 6 | | Intentional($DS_{16}$) | 30 | 11410 | 918 | 2.74 | 127 |
| 7 | | Intentional($DS_{32}$) | 23 | 10200 | 368 | 1.4 | 41 |
| 8 | | Natural & Intentional($DS_2$) | 49 | 83280 | 1453 | 2.82 | 204 |
| 9 | | Natural & Intentional($DS_4$) | 42 | 27921 | 1127 | 2.88 | 165 |
| 10 | | Natural & Intentional($DS_8$) | 39 | 13475 | 913 | 2.81 | 132 |
| 11 | | Natural & Intentional($DS_{16}$) | 30 | 10647 | 653 | 2.27 | 7 |

*Number of Gate Equivalents, **Nanoseconds

Table 4 Cost-accuracy trade-off in conventional and some selected BPC implementations of the FRNN hardware (absolute implementation result).

| Row# | Realization Type | Utilized Sparsity Type | FRNN Accuracy Simulation Results | | | Single Neuron Implementation Results | | | |
|---|---|---|---|---|---|---|---|---|---|
| | | | CCR | TE | MSE | Two-Level (# of literals) | Area (# of GE*) | Delay (ns**) | Power (uw***) |
| 1 | Conventional | - | 89 | 170 | 0.026 | 330093 | 1432 | 3.57 | 245 |
| 2 | Bioinspired Partially Precise Computing (BPC) | Natural | 89 | 170 | 0.026 | 206231 | 1715 | 4.49 | 253 |
| 3 | | Intentional ($TH_{48}^{48}$) | 87 | 240 | 0.032 | 291184 | 2142 | 5.21 | 300 |
| 4 | | Intentional ($DS_{16}$) | 85 | 330 | 0.037 | 6227 | 617 | 3.51 | 85 |
| 5 | | Intentional ($DS_{32}$) | 85 | 600 | 0.048 | 5735 | 345 | 2.05 | 44 |
| 6 | | Natural & Intentional ($DS_{16}$) | 85 | 330 | 0.037 | 5981 | 548 | 2.9 | 67 |
| 7 | | Natural & Intentional ($DS_{32}$) | 85 | 600 | 0.048 | 5699 | 317 | 1.94 | 39 |
| 8 | | Natural & Intentional ($TH_{48}^{48}$ + $DS_{16}$) | 82 | 250 | 0.040 | 5940 | 586 | 3.16 | 74 |
| 9 | | Natural & Intentional ($TH_{48}^{48}$ + $DS_{32}$) | 85 | 250 | 0.052 | 5688 | 315 | 1.87 | 40 |

*Number of Gate Equivalents, **Nanoseconds, ***Micro Watt